\documentclass[twocolumn,prl,aps,superscriptaddress]{revtex4-2}

\usepackage[latin9]{inputenc}
\usepackage{mathtools}
\usepackage{amsbsy}
\usepackage{amssymb}
\usepackage{amstext}
\usepackage{bm}
\usepackage{xcolor}
\usepackage{cancel}
\definecolor{darkblue}{rgb}{0, 0, 1}
\newcommand{\RN}[1]{%
	\textup{\uppercase\expandafter{\romannumeral#1}}%
}
\usepackage[unicode=true,pdfusetitle,
bookmarks=false,
breaklinks=false,pdfborder={0 0 1},backref=false,colorlinks=true,allcolors=darkblue]
{hyperref}
\hypersetup{
	bookmarksnumbered=false,bookmarksopen=false}
\makeatletter

\@ifundefined{textcolor}{}
{%
	\definecolor{BLACK}{gray}{0}
	\definecolor{WHITE}{gray}{1}
	\definecolor{RED}{rgb}{1,0,0}
	\definecolor{GREEN}{rgb}{0,1,0}
	\definecolor{BLUE}{rgb}{0,0,1}
	\definecolor{CYAN}{cmyk}{1,0,0,0}
	\definecolor{MAGENTA}{cmyk}{0,1,0,0}
	\definecolor{YELLOW}{cmyk}{0,0,1,0}
}

\newcommand{\beq}{\begin{equation}}
\newcommand{\eeq}{\end{equation}}
\newcommand{\beqa}{\begin{eqnarray}}
\newcommand{\eeqa}{\end{eqnarray}}

\usepackage{amsmath}
\usepackage{amssymb}
\usepackage{txfonts,color}
\makeatother

\begin{document}
\title{Phase-Programmable Free Electron Quantum States in Synthetic Momentum Space}

\author{Alatz Alvarez-Ahedo}

\affiliation{Department of Physical Chemistry, University of the Basque Country UPV/EHU, Apartado 644, 48080 Bilbao, Spain}

\author{Miriam Lazo}
\affiliation{Department of Physical Chemistry, University of the Basque Country UPV/EHU, Apartado 644, 48080 Bilbao, Spain}

\author{Tian-Niu Xu}
\affiliation{College of Information Science and Engineering, Qilu Normal University, Jinan 250013, China}

\author{Yiming Pan}
\affiliation{State Key Laboratory of Quantum Functional Materials, School of Physical Science and Technology and Center for Transformative Science, ShanghaiTech University, Shanghai 200031, China}

\author{Mikel Sanz}
\affiliation{Department of Physical Chemistry, University of the Basque Country UPV/EHU, Apartado 644, 48080 Bilbao, Spain}

\author{Yongcheng Ding}
\affiliation{Department of Physical Chemistry, University of the Basque Country UPV/EHU, Apartado 644, 48080 Bilbao, Spain}

\date{\today}
\begin{abstract}
Light-electron interactions generate synthetic momentum-space dynamics that can be used to engineer free electron quantum states. Here we develop coherent control protocols in which the optical phase acts as the controllable hopping phase of a Floquet-Bloch momentum lattice. Pontryagin optimization designs phase-only waveforms that prepare selected momentum populations and coherent few-sideband superpositions with programmable relative phases. In a complementary Bragg regime protocol, dynamical phase matching selectively couples neighboring sidebands and enables deterministic sequential state synthesis. Full wave-packet simulations based on the minimal-coupling Hamiltonian identify the tolerance window set by phase noise, detuning, and finite momentum spread. The two protocols expose a speed-selectivity tradeoff between ultrafast multilevel interference control and slower resonant engineering, establishing programmable free electron sidebands as a platform for ultrafast quantum state synthesis.
\end{abstract}

\maketitle

\section{Introduction}

Free electrons constitute one of the most fundamental quantum platforms for probing and manipulating matter beyond conventional bound state systems. Recent advances in ultrafast electron science have enabled precise control of electron beams in both space and time~\cite{hemsing2014beam,petrillo2013observation,abajo2023spatiotemporal,mihaila2022transverse,velasco2025free}, driving major developments in electron microscopy~\cite{barwick2009photon,park2010photon,abajo2010multiphoton,feist2015quantum,dahan2021imprinting}, diffraction~\cite{jones2016laser,eldar2024self,pan2019anomalous,lin2024ultrafast,ding2025ultrafast}, coherent radiation~\cite{ayvazyan2002generation,grgyas2012ultrafast}, and free electron laser~\cite{robles2025spectrotemporal,hu2025tunable}. As electron pulses approach the femtosecond and attosecond regimes~\cite{aseyev2020the}, the coherent structure of the electron wavefunction itself becomes an essential physical resource. The ability to engineer free electron quantum states enables direct control over quantum interference~\cite{turner2021interaction,johnson2022inelastic,echternkamp2016ramsey,feist2020high,bucher2023}, light-matter interaction~\cite{karnieli2023jaynes,synanidis2024quantum,karnieli2024universal,ding2025gates}, and nonequilibrium electron dynamics~\cite{cairano2026nonequilibrium}, establishing the foundation of free electron quantum optics~\cite{ruimy2025free}. Beyond ultrafast imaging~\cite{priebe2017attosecond,asban2021generation,bucher2024coherently,gaida2024attosecond} and sensing~\cite{ruimy2021toward,gorlach2021ultrafast,karnieli2023quantum,velasco2025quantum}, coherent wavefunction shaping further opens opportunities for synthetic quantum dynamics~\cite{braiman2023the,pan2024synthetic,chen2026synthetic} and quantum simulation~\cite{pan2024free,sirotin2024quantum} with electron beams.

A particularly intriguing feature of coherent light-electron interactions is the emergence of synthetic momentum space dynamics. Due to the spatiotemporal periodicity of the driving electromagnetic field, Floquet-Bloch theory maps the free electron wavefunction onto a discrete set of momentum sidebands separated by integer photon quanta. Optical absorption and emission processes coherently couple neighboring sidebands, generating an effective tight binding dynamics in synthetic momentum space. This correspondence connects free electron quantum optics with a broad range of lattice phenomena, including Bloch oscillations~\cite{battesti2004bloch,stockhofe2015bloch,ali2024floquet}, Wannier-Stark dynamics~\cite{glueck2002wannier,zhang2025formation,dikopoltsev2025collective}, and synthetic fields~\cite{lin2011a,zapletal2019dynamically,tavakol2021artificial,zhang2025synthetic}. At the same time, the momentum sideband amplitudes and relative phases directly encode the spatiotemporal structure of the electron wavefunction, suggesting the possibility of programmable quantum state engineering through controlled synthetic lattice dynamics.

However, existing transport mechanisms in synthetic momentum space remain largely restricted to specific dynamical processes governed by adiabatic evolution or simple resonant transitions. While such mechanisms provide valuable physical intuition, they generally do not offer a systematic route toward arbitrary coherent state preparation with independently programmable amplitudes and phases. This difficulty becomes particularly severe in multilevel diffraction regimes, where many momentum sidebands participate simultaneously in the evolution and generate increasingly complicated interference dynamics. In addition, coherent control for low energy electrons may require experimentally demanding optical intensities in order to enter the desired diffraction regime. Developing fast and programmable protocols for coherent free electron quantum state engineering therefore remains an important challenge.

Quantum optimal control provides a natural framework for addressing this problem. Rather than relying on adiabatic control or going for efficient analytical or numerical protocols~\cite{torosov2011smooth,genov2014correction,odelin2019shortcuts,ding2021breaking,takahashi2024shortcuts}, numerical optimal control~\cite{khaneja2005optimal,reich2012monotonically,ding2020smooth} directly searches for control fields that maximize a target objective functional under constraints on experiments without requiring explicit analytical solutions or prohibitive computational resources. Over the past decades, optimal control theory~\cite{pontryagin2018mathemagtical,boscain2021introduction} has enabled high fidelity quantum manipulation across a wide range of physical systems and tasks, including cold atoms~\cite{hocker2016optimal,sorensen2018quantum,larrouy2020fast,dupont2021quantum}, cavity quantum electrodynamics~\cite{heeres2017implementing,lewalle2023pontryagin,karmakar2026cdj}, trapped ions~\cite{chen2011optimal,fuerst2014controlling,evangelakos2023minimum}, superconducting circuits~\cite{zhou2024optimal,dong2026pontryagin}, and even quantum algorithms~\cite{yang2017optimizing}. In the context of synthetic momentum dynamics, optimal control offers a particularly attractive route toward preparing arbitrary free electron superposition states beyond conventional adiabatic protocols.

On the other hand, the Bragg regime reveals an alternative control mechanism rooted in the intrinsic frequency selectivity of the synthetic momentum lattice~\cite{pan2023low}. In this regime, neighboring momentum sidebands can be individually addressed through dynamical phase matching, allowing coherent population transfer to proceed through effectively resonant two-level dynamics. This observation establishes a direct analogy between free electron momentum sidebands and ladder state control protocols developed in cavity quantum electrodynamics. Inspired by the Law-Eberly mechanism~\cite{law1996arbitrary}, one may therefore construct deterministic state preparation protocols based on sequential resonant coupling between neighboring sidebands.

In this work we combine these two ideas. Starting from the minimal-coupling Hamiltonian, we derive an effective Floquet-Bloch lattice model and formulate phase-only Pontryagin optimization for free-electron state preparation. We then demonstrate preparation of representative sideband populations, momentum-space cat states, and three-sideband coherent superpositions using full wave-packet simulations for validation. Finally, we introduce a deterministic Bragg-regime protocol based on frequency-selective sideband-pair rotations. The resulting picture is close to the original synthetic-momentum-space manuscript, but the central claim is sharpened: fast global interference control and slow resonant selectivity are complementary limits of programmable free-electron quantum-state synthesis.

\section{Floquet-Bloch Model and Single Parameter Control}
\label{sec:fb}
We begin with the Hamiltonian of a relativistic free electron interacting with an electromagnetic field through minimal coupling, and expand the kinetic energy to second order around the central momentum,
\begin{equation}
\label{eq:tdse}
H = E_0 + v_0(p-p_0) + \frac{(p-p_0)^2}{2\gamma^3m_e} - \frac{e}{2\gamma m_e}(A\cdot p + p\cdot A),
\end{equation}
where $p_0 = \gamma m_e v_0$ is the central momentum of the electron and $A$ is the vector potential. In the presence of the optical grating, the vector potential takes the form
\begin{equation}
A(z,t) = \frac{E_z}{\omega_L}\cos(\omega_L t -qz+\phi),
\end{equation}
with laser frequency $\omega_L$, field amplitude $E_z$, wave vector $q=2\pi/\Lambda$, and grating period $\Lambda$, where the phase-matching condition is given by $v_0q=\omega_L$. We employ Floquet-Bloch theory and expand the electron wavefunction as
\begin{equation}
\Psi(z,t) = \sum_nc_n(t)e^{-i(\omega_nt-k_nz)},
\end{equation}
where $\omega_n=E_0/\hbar + n\omega_L$ and $k_n=k_0+nq$ follow from the spatiotemporal periodicity of the interaction. Under the non-recoil approximation, $k_n\approx k_0$, the time-dependent Schr\"odinger equation reduces to the coupled-mode equation
\begin{equation}
\label{eq:tbeq}
i\dot{c_n} = n^2\epsilon c_n + \kappa c_{n+1} + \kappa^* c_{n-1}.
\end{equation}
Here, $\epsilon=\frac{\hbar q^2}{2\gamma^3m_e}$ denotes the on-site energy associated with the relativistic band curvature, while $\kappa=-\frac{ek_0E_z}{2\gamma m_e\omega_L}e^{i(\phi+\pi/2)}$ is the hopping amplitude. Equation~\eqref{eq:tbeq} describes the electron dynamics in both the Bragg and Raman-Nath regimes, corresponding respectively to weak-field or slow-electron interactions, and strong-field or fast-electron interactions. The relative importance of the on-site and hopping terms is characterized by the Klein-Cook parameter $Q=\epsilon/2|\kappa|$. In the Bragg regime ($Q\gg1$), the on-site term dominates and suppresses coupling to higher order sidebands.

\begin{figure}
\includegraphics[width=1\linewidth]{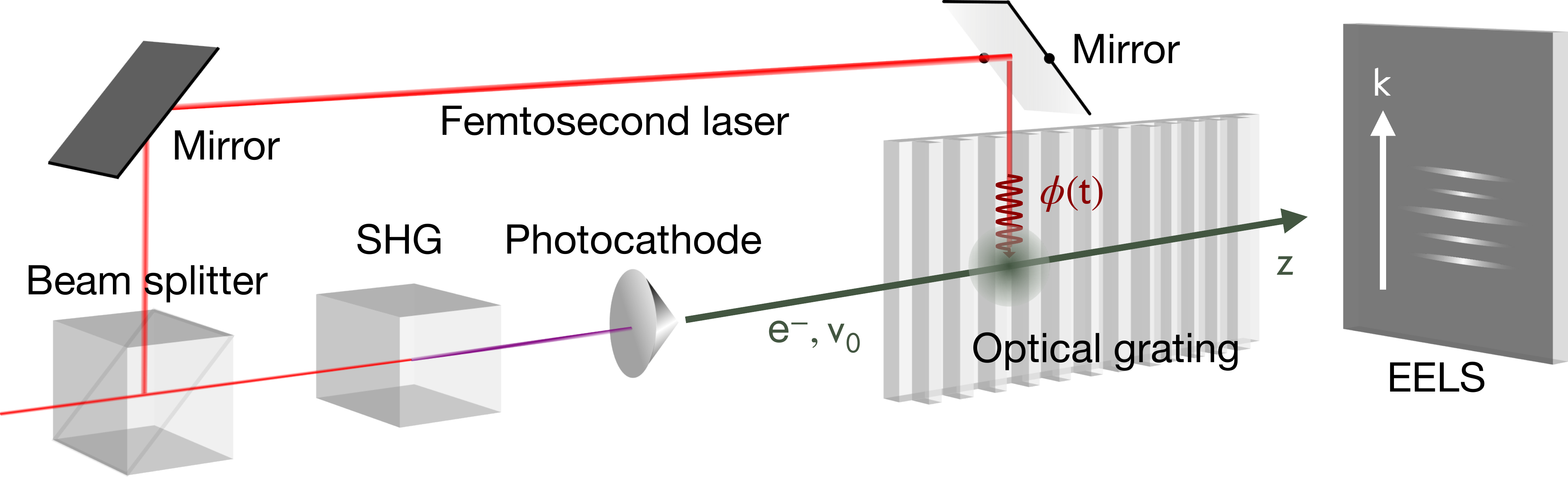}
\caption{\label{fig:scheme} Schematic of the experimental setup. A femtosecond laser pulse is split into two branches. One branch generates free electrons from a photocathode through second harmonic excitation, producing an electron beam with central velocity $v_0$. The second branch drives the light-electron interaction at the optical grating with a programmable phase $\phi(t)$. The resulting phase-controlled diffraction generates tunable momentum spectra measured by high resolution electron energy-loss spectroscopy.}
\end{figure}

The experimental setup is shown in Fig.~\ref{fig:scheme}. Free electrons are emitted from a photocathode driven by one branch of a femtosecond laser pulse, following second-harmonic generation to produce ultraviolet excitation. The emitted electron wavepacket carries a central momentum $p_0$, corresponding to the initial condition $c_0(0)=1$. The second laser branch is shaped by a spatial light modulator to generate a programmable phase profile $\theta(z,t)$, enabling the realization of topological bound states and electron Stern-Gerlach dynamics. An optical grating maintains the phase matching condition throughout the interaction region. In the present work, we focus on single parameter control, namely the manipulation of the electron momentum spectrum through a time-dependent phase modulation $\theta(t)$. After passing through the interaction region, the electron undergoes momentum space diffraction, and the resulting momentum spectrum can be measured using electron energy-loss spectroscopy.

For simplicity, we rewrite the hopping amplitude as $\kappa\rightarrow|\kappa|e^{i\phi(t)}$ in~\eqref{eq:tbeq}, such that the dynamics can be expressed in matrix form as
\begin{equation}
i\dot{C}=M[\phi(t)]C.
\end{equation}
Given an initial condition, the dynamics define an optimal control problem aimed at maximizing an objective functional $J[C(T),C^\dag(T)]$ within a fixed operation time $T$. The specific form of the objective functional depends on the target task. Because the control enters through a nonlinear phase-dependent hopping term, we adopt Pontryagin's maximum principle, which naturally accommodates the single parameter control structure.

The Pontryagin Hamiltonian is given by
\begin{equation}
H_P = \text{Re}[\langle\tilde{C}|\dot{C}\rangle] = \text{Im}[\langle \tilde{C}|M|C\rangle],
\end{equation}
where $\tilde{C}$ denotes the costate vector satisfying the final time boundary condition
\begin{equation}
\tilde{C}(T)=\partial J[C(T),C^\dag(T)]/\partial C^\dag(T).
\end{equation}
The optimal control field $\phi^*(t)$ maximizes the Pontryagin Hamiltonian at each instant of time,
\begin{equation}
\phi^*(t)=\arg\max_{\phi(t)}H_P.
\end{equation}
In general, obtaining the exact maximizer is analytically intractable for the present control problem. We therefore implement a numerical realization of Pontryagin's maximum principle through iterative gradient ascent toward a stationary solution satisfying $\partial H_P/\partial\phi(t)=0$. The control update is determined from the functional derivative of $H_P$ with respect to $\phi(t)$:
\begin{equation}
\frac{\partial H_P}{\partial\phi(t)}
=
\text{Im}\left[\langle \tilde{C}|\frac{\partial M}{\partial\phi(t)}|C\rangle\right].
\end{equation}

The optimization procedure is summarized as follows:

(i) Initialize the control field $\phi(t)$ from a chosen ansatz.

(ii) Propagate the state $C$ forward in time and the costate $\tilde{C}$ backward in time using the corresponding initial and final boundary conditions.

(iii) Update the control field according to the learning rate $\alpha$ via $\phi(t)\rightarrow\phi(t)+\alpha\frac{\partial H_P}{\partial\phi(t)}.$

(iv) Repeat step (ii) until the convergence criterion is satisfied.

\section{Momentum State Population Engineering}
\label{sec:pop}
We first investigate the control of the sideband populations $P_n=|c_n|^2$ in momentum space. Using the experimental setup shown in Fig.~\eqref{fig:scheme}, we initialize the electron in the central momentum state with $c_0(0)=1$, corresponding to the phase matching condition $v_0q=\omega_L$. Our objective is to engineer the electron momentum spectrum such that the final sideband populations approach a target distribution $P_{tn}$. Accordingly, the cost functional $J[C(T),C^\dag(T)]$ is chosen as the classical fidelity
\begin{equation}
J[C(T),C^\dag(T)] = F_{\text{C}} = \left[ \sum_n {\sqrt{P_nP_{tn}}} \right]^2,
\end{equation}
where the final time boundary condition for the costate is
\begin{equation}
\tilde{C}(T) = \frac{\partial F_{\text{C}} }{\partial C^\dag(T)} = \sqrt{F_C P_n} \dfrac{c_n}{|c_n|}.
\end{equation}

Using this framework, we first demonstrate ultrafast preparation of a single sideband state, followed by the generation of more complex momentum distributions. Population-based control is particularly relevant in free electron experiments where the measured observable is the final energy or momentum spectrum rather than the coherent wavefunction itself. Typical examples include electron energy-gain spectroscopy and momentum-resolved beam shaping, where detector signals depend primarily on the sideband occupation probabilities. In such scenarios, the relative phases between different sidebands do not directly affect the measured outcome and are therefore excluded from the optimization target. Nevertheless, coherent light-electron interaction remains essential throughout the control process, since the final population transfer still originates from quantum interference during the dynamical evolution.

\subsection{Single Momentum State}
We first consider the optimization of the control field $\phi(t)$ for the preparation of a single target sideband $|n\rangle$. This process is closely related to the acceleration and deceleration of free electrons in dielectric laser acceleration (DLA). In conventional DLA theory, however, electrons are typically treated as classical point particles, whereas the present framework describes the electron as a coherent matter wave through the Floquet-Bloch theory underlying the coupled-mode dynamics.
In practice, the electron cannot be represented by an ideal Floquet-Bloch plane-wave ansatz and instead remains spatially localized. To account for this effect, we model the initial electron state as a Gaussian wavepacket in momentum space,
\begin{equation}
|n\rangle = \mathcal{N}\exp\left[-\frac{(k-k_0-nq)^2}{2\sigma_k^2}\right],
\end{equation}
where we choose $n=0$ and $\sigma_k=0.05q$ for initialization. Using the optimized control field $\phi^*(t)$, we then simulate the full time-dependent Schr\"odinger equation governed by the minimal-coupling Hamiltonian~\eqref{eq:tdse} using the Crank-Nicolson method, yielding the final momentum space wavefunction $\Psi_k(T)$. The sideband populations are subsequently extracted by integrating the momentum distribution around each sideband,
\begin{equation}
P_n = \int_{k_0+nq-q/2}^{k_0+nq+q/2}|\Psi_k(T)|^2dk,
\end{equation}
which allows us to evaluate the experimental fidelity $F_{\text{TDSE}}$ beyond the ideal coupled-mode approximation.

In the numerical simulations, we choose the dimensionless on-site energy $\epsilon =1$, hopping amplitude $|\kappa|=1.25$, and a fixed operation time $T=4$. The corresponding Klein-Cook parameter is $Q=0.4$, placing the interaction in the intermediate regime between the Bragg and Raman-Nath limits. The optimization for $|1\rangle$ and $|2\rangle$ is initialized with a quadratic ansatz $\phi(t)\propto t^2$ inspired by Bloch acceleration. For $|3\rangle$, we use a hybrid ansatz combining a smooth hyperbolic-tangent phase step with a sinusoidal modulation. For $|4\rangle$, the optimization is initialized with a piecewise chirped phase, constructed by accumulating successive resonant frequency steps. These ansatze are chosen empirically based on Bloch acceleration and bang-bang like solution in optimal control theory.  After sufficient iterations under learning rate $\alpha=0.05$, the optimal control field $\phi^*(t)$ is obtained by selecting the solution with the highest achieved fidelity $F^*$. 

Considering mapping to laboratory parameters, the electron momentum energy is set to be 640 eV, corresponding to group velocity of $v_g=0.05c$. The optical field has a photon energy of $\hbar\omega_L = 6.2~\mathrm{eV}$ in the ultraviolet regime of wavelength $\lambda = 200\mathrm{nm}$. Therefore, for phase matching of the light-electron interaction, we need to engrave the optical grating of periodicity $\Lambda=10\mathrm{nm}$. Accordingly, the on-site amplitude and hopping amplitude give $\epsilon=2.275\times10^{-2}~1/\mathrm{fs}$ and $|\kappa|=2.844\times10^{-2}~1/\mathrm{fs}$, requiring a field amplitude $E_z=2.35\times 10^7~\mathrm{V/m}$. realizing the quantum control within $T=175.81~\mathrm{fs}$. These parameters are not changed unless specifically mentioned.

\begin{figure}
\includegraphics[width=1\linewidth]{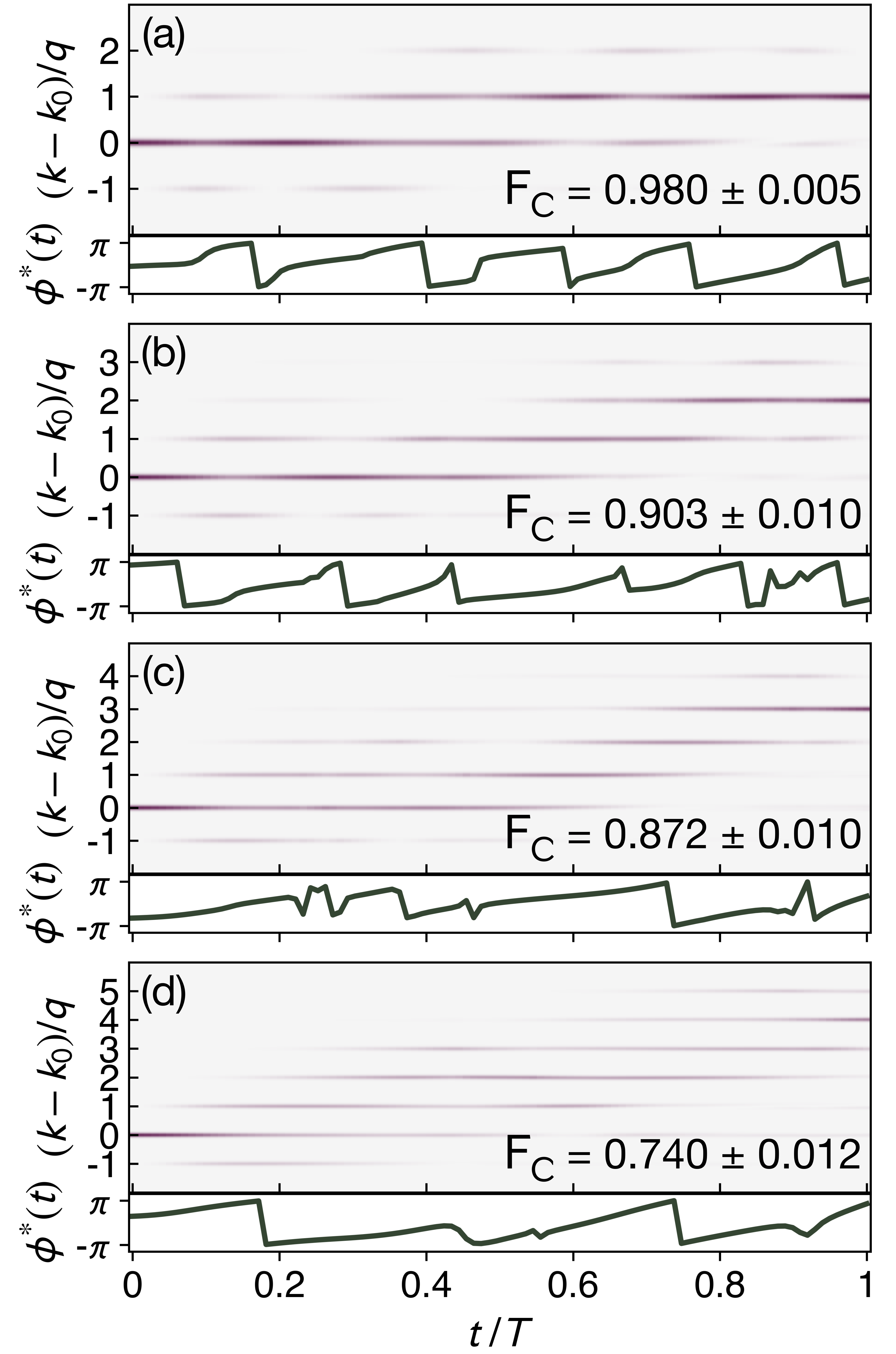}
\caption{\label{fig:single} Single-momentum state preparation through single parameter control. The optimized control field $\phi^*(t)$ and corresponding momentum space dynamics are shown for target states (a) $|c_1(T)|^2 =1$, (b) $|c_2(T)|^2 =1$, (c) $|c_3(T)|^2 =1$, and (d) $|c_4(T)|^2 =1$, initialized from $c_0(0)=1$. Simulations are performed with dimensionless hopping amplitude $|\kappa|=1.25$, on-site energy $\epsilon=1$, and operation time $T=4$. The averaged classical fidelities are obtained from 10 realizations with Gaussian phase noise applied to the control field, which the errors denote 95\% confidence intervals.}
\end{figure}

Figure~\ref{fig:single} presents the optimized transfer from the initial state to target sidebands ranging from $n=1$ to $n=4$, together with the corresponding momentum space dynamics and optimal control fields. Owing to the periodicity of the phase variable, the optimized phase profiles are wrapped within the interval $[-\pi,\pi]$. We further examine the robustness of the control protocol by introducing Gaussian control noise according to
\begin{equation}
\phi^*(t)\rightarrow\phi^*(t)+\mathcal{G}(\mu=0,\sigma=0.05\pi).
\end{equation}
The experimentally simulated fidelities $F_C$, averaged over 10 noise realizations, remain in good agreement with the ideal optimization fidelities $F^*_C=\{0.991,0.921,0.894,0.756\}$. Both fidelities gradually decrease as the target sideband index increases, reflecting the increasing complexity of high-order population transfer. Efficient preparation of substantially higher sidebands would require longer operation times and a more demanding optimization landscape. 

We further note that single-sideband preparation can alternatively be interpreted as a form of Bloch-like acceleration in synthetic momentum space. Under the gauge transformation $c_n(t)=a_n(t)e^{-in\phi(t)}$, Eq.~\eqref{eq:tbeq} becomes
\begin{equation}
i\dot{a}_n = \left[n^2\epsilon - n\dot{\phi}(t)\right]a_n + |\kappa|\left(a_{n+1}+a_{n-1}\right),
\end{equation}
where the time-dependent phase modulation generates an effective linear potential proportional to $\dot{\phi}(t)$. The center of the resulting quadratic potential is therefore shifted to
$n_c={\dot{\phi}(t)}/{2\epsilon}$. The corresponding adiabatic condition is generally more complicated than the harmonic-oscillator criterion obtained from the broad wavepacket and continuum approximations, $|\ddot{\phi}(t)|\ll\mathcal{O}(\epsilon^{5/4}|\kappa|^{3/4})$. In the Bragg regime $\epsilon\gg|\kappa|$, the dynamics instead become Landau-Zener-like, yielding the approximate condition $|\ddot{\phi}(t)|\ll|\kappa|^2$. In the hybrid diffraction regime $|\kappa|\gtrsim\epsilon$, the adiabatic criterion becomes more involved and approximately scales as $|\ddot{\phi}(t)|\ll\epsilon|\kappa|$. These results indicate that the optimal control dynamics interpolate between adiabatic Bloch-like transport and strongly nonadiabatic multilevel diffraction as the Klein-Cook parameter is varied.

\subsection{Programmable Momentum Distributions}

In the second set of numerical experiments, we investigate the capability of the optimal control protocol to prepare arbitrary momentum space population distributions. Unlike the single-sideband preparation discussed previously, these target states involve simultaneous control over multiple momentum channels and therefore provide a more stringent test of the optimization framework.

\begin{figure}
\includegraphics[width=1\linewidth]{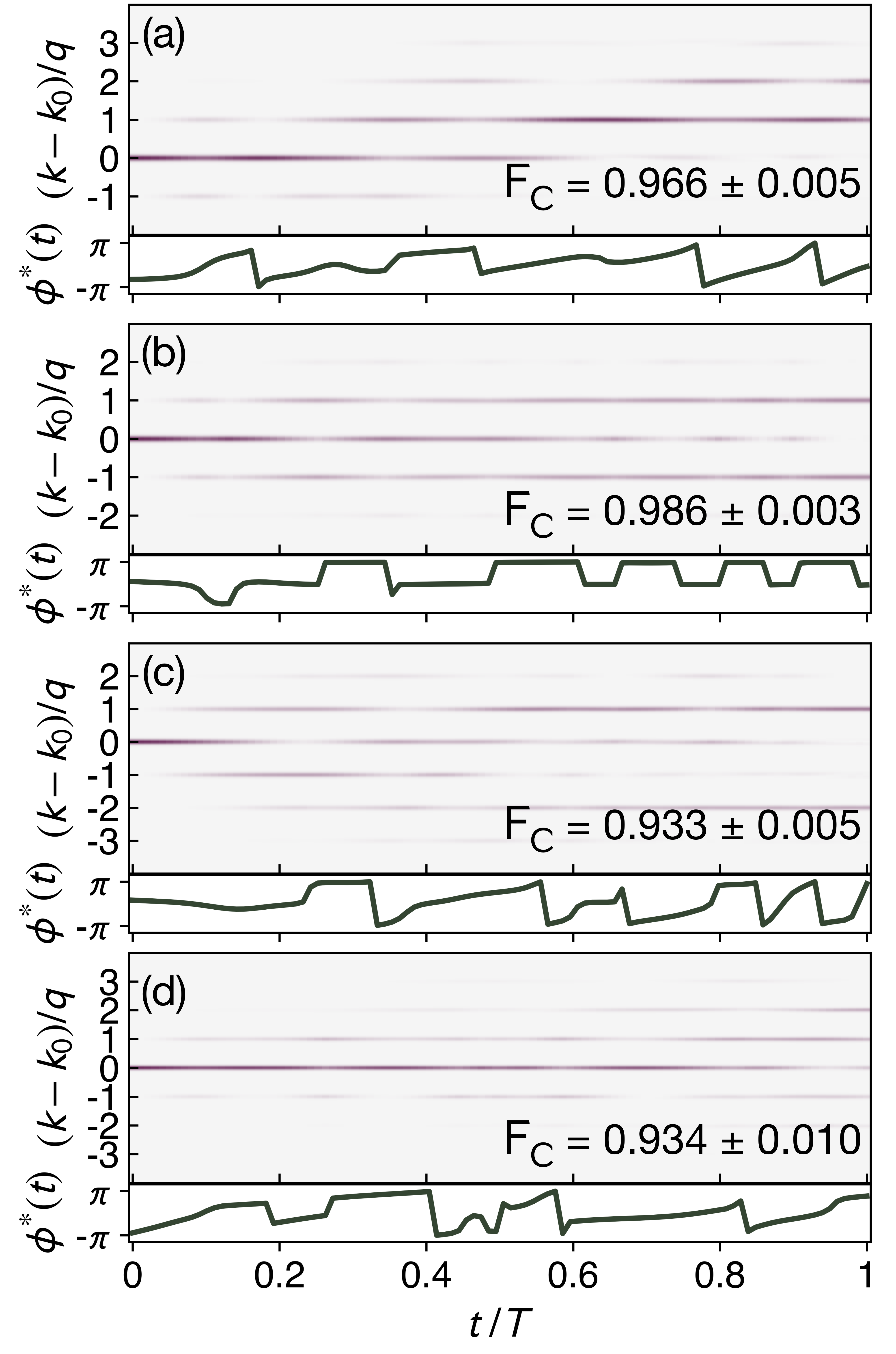}
\caption{\label{fig:multi} Arbitrary momentum-population preparation through single parameter control. Target population distributions: (a) $P_1:P_2=1:1$, (b) $P_{-1}:P_{1}=1:1$, (c) $P_{-2}:P_1=1:1$, and (d) $P_{-2}:P_{-1}:P_0:P_1:P_2=1:2:3:4:5$. Same parameters and definitions as in Fig.~\ref{fig:single}.}
\end{figure}

We first consider equal-weight population transfer into selected sidebands, including neighboring sidebands [Fig.~\ref{fig:multi}(a)], symmetric opposite sidebands [Fig.~\ref{fig:multi}(b)], and nonuniform arbitrary sideband configurations [Fig.~\ref{fig:multi}(c)]. We then extend the protocol to more complex momentum distributions involving larger numbers of populated sidebands with independently chosen target weights [Fig.~\ref{fig:multi}(d)]. These examples demonstrate the flexibility of the control protocol in engineering programmable free electron spectra beyond simple resonant transitions.

For all tested configurations, the numerical optimization achieves high classical fidelities $F_C^*=\{0.972,0.991,0.944,0.994\}$. To evaluate the robustness of the optimized control fields, we additionally introduce Gaussian perturbations to the ideal phase modulation and simulate the corresponding dynamics under the full time-dependent Schr\"odinger equation. The resulting fidelities remain above $90\%$ for all demonstrated cases, indicating that the optimized protocols remain stable against moderate systematic phase errors.

The ability to engineer arbitrary momentum space population distributions also suggests a route toward programmable spectral synthesis with free electrons. In analogy to programmable momentum state preparation in ultracold atomic systems, the populated sidebands may be interpreted as configurable spectral channels generated through controlled light-electron interactions. By combining optimized optical phase modulation with energy-resolved electron detection, one may build a "free electron printer'' capable of producing programmable momentum space patterns at ultrafast timescales. Such controllable spectral shaping could provide useful functionalities for electron beam engineering, ultrafast spectroscopy, and dynamically programmable free electron sources.

\section{Coherent momentum state Engineering}
\label{sec:coherent}
In Section~\ref{sec:pop}, we optimized the control field $\phi(t)$ toward target states defined solely by the sideband populations. Here, we extend the optimization to coherent quantum state preparation, where both the populations and the relative phases between momentum sidebands are controlled simultaneously. The cost functional is therefore chosen as the quantum fidelity
\begin{equation}
J[C(T),C^\dag(T)] = F_{\text{Q}} = |\langle C(T)|C_t\rangle|^2,
\end{equation}
where $|C_t\rangle$ denotes the target wavefunction. The final time boundary condition for the costate is given by
\begin{equation}
\tilde{C}(T) = \frac{\partial F_{\text{Q}} }{\partial C^\dag(T)} = [\langle C(T)|C_t\rangle]C_t.
\end{equation}

Precise control of the relative phases between sidebands is essential for coherent free electron manipulation, since interference between momentum components determines the real space and temporal structure of the electron wavepacket. Such phase-engineered states are directly relevant for attosecond electron pulse shaping, coherent beam splitting, quantum interferometry, and quantum information encoding in synthetic momentum space lattices. We first demonstrate the preparation of a two-component superposition state with programmable relative phase, followed by the preparation of more complex multi-component states. In addition, we investigate the influence of imperfect phase matching and increased spatial localization of the initial electron wavepacket.

\subsection{Momentum-space Cat States}

\begin{figure}
\includegraphics[width=1\linewidth]{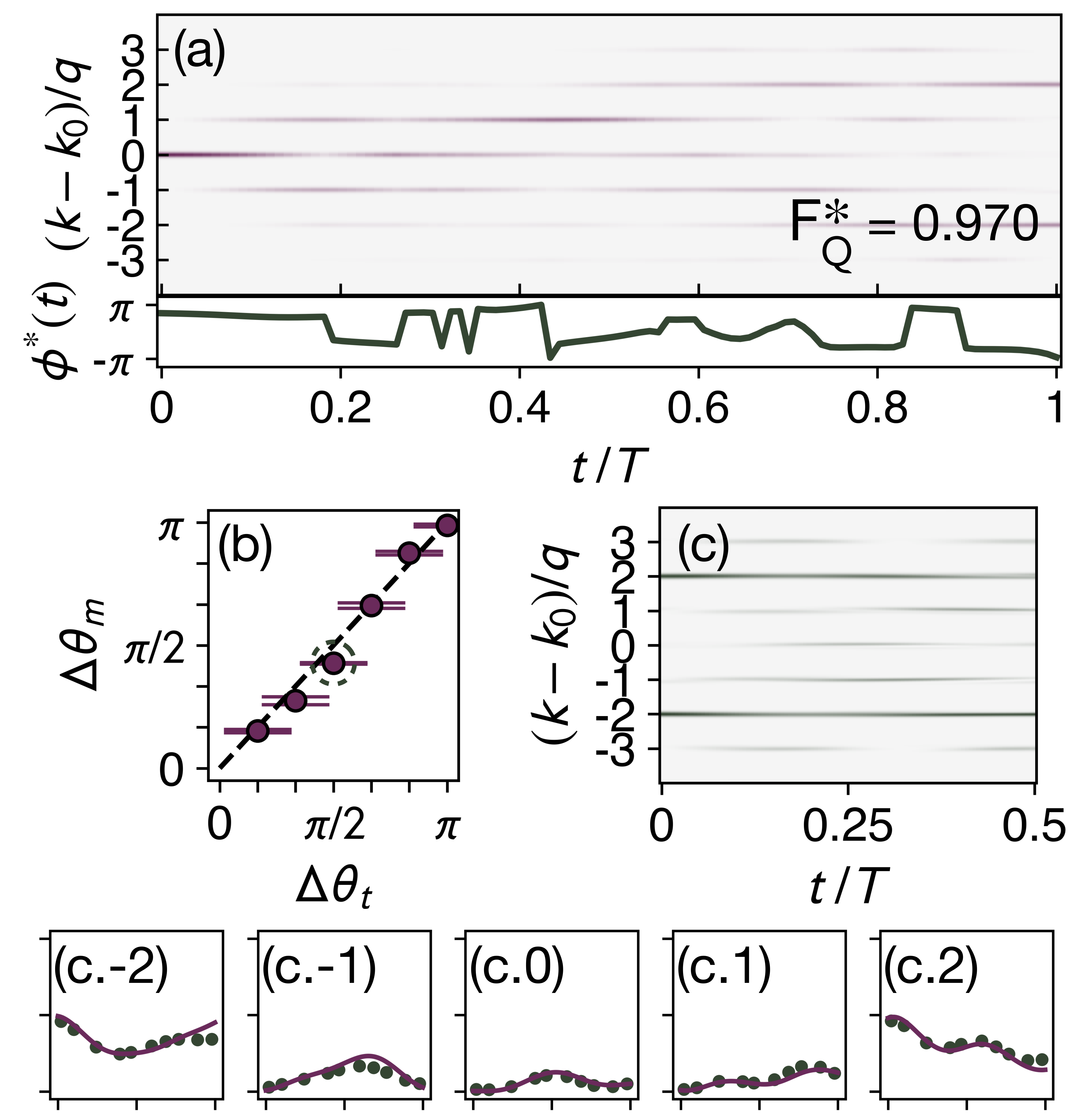}
\caption{\label{fig:cat} Cat-state preparation through single parameter control for the target state $|C_t\rangle=\frac{1}{\sqrt{2}}\left(|-2\rangle + e^{i\Delta\theta}|2\rangle\right).$ (a) Optimized control field $\phi(t)$ and corresponding momentum space dynamics for the target relative phase $\Delta\theta=\pi/2$. Pontryagin's optimization yields a quantum fidelity of $F_Q^*=0.97$ with Floquet-Bloch theory. (b) Experimentally reconstructed relative phases $\Delta\theta_m$ compared with the target phases $\Delta\theta$. (c) Interferometric phase reconstruction for the circled case in (a) and (b). After the control sequence, the electron undergoes additional free evolution for a duration of $0.5T$ under a static control field $\phi(t)=0$. The measured sideband populations are fitted to numerical simulations generated from Gaussian sideband ansatz states with different relative phase perturbations. Panels (c.-2 to 2) show the corresponding least square fitting for each sideband. Error bars denote the 95\% confidence intervals. Other parameters are not changed.}
\end{figure}

To illustrate coherent quantum control, we prepare target states consisting of superpositions of two symmetric momentum sidebands,
\begin{equation}
|C_t\rangle = \frac{1}{\sqrt{2}}\left(|-2\rangle + e^{i\Delta\theta}|2\rangle\right),
\end{equation}
corresponding to momentum space cat states with tunable relative phase $\Delta\theta$. We prepare a family of cat states with target phases $\Delta\theta_j = j\pi/6$, where $j\in\{1,2,3,4,5,6\}$. After obtaining the optimal control field $\phi^*(t)$, we simulate the full time-dependent Schr\"odinger equation to evaluate the control fidelity beyond the ideal coupled-mode approximation. Figure~\ref{fig:cat}(a) shows the momentum space dynamics for the target phase $\Delta\theta_t=\pi/2$ driven by the optimized control field, corresponding to the circled case in Fig.~\ref{fig:cat}(b).

Unlike population-only control, verification of a quantum state additionally requires extraction of the relative phase between sidebands from the final wavefunction $\Psi_k(T)$. In free electron experiments, a feasible measurement protocol is to further evolve the electron under a static control field $\phi(t)=0$ and monitor the subsequent sideband population oscillations using electron energy-loss spectroscopy, analogous to interferometric phase reconstruction [see Fig.~\ref{fig:cat}(c)]. Meanwhile, the coupled-mode optimization yields the final Floquet-Bloch state $|C(T)\rangle$ with ideal quantum fidelity $F_Q^*$. We therefore reconstruct an initial state consisting of a coherent superposition of Gaussian sidebands,
\begin{eqnarray}
\label{eq:gaussian_cat}
|\Psi\rangle &=& \mathcal{N}\left\{\sum_{n\neq2} c_i \exp\left[-\frac{(k-k_0-nq)^2}{2\sigma_k^2}\right] \right.\nonumber\\
&+& \left. c_2 e^{i\delta\theta}\exp\left[-\frac{(k-k_0-2q)^2}{2\sigma_k^2}\right]\right\},
\end{eqnarray}
where $\delta\theta$ represents an additional phase perturbation applied to the $|2\rangle$ sideband. Different perturbed initial states are subsequently evolved under the static control field $\phi(t)=0$ to generate reference population dynamics. The experimentally measured sideband populations are then fitted to the reference evolution through least-square minimization over $\delta\theta$ [see Fig.~\ref{fig:cat}(c.-2 to 2)], yielding the reconstructed relative phase
\begin{equation}
\Delta\theta_m = \text{arg}(c_2/c_{-2}) + \delta_\theta^*.
\end{equation}

After reconstructing the coherent superposition from the TDSE wavefunction $\Psi_k(T)$, we evaluate the quantum fidelity relative to the target Floquet-Bloch state. The resulting averaged fidelities are $F_Q=\{0.918,0.914,0.911,0.910,0.910,0.914\}$ for all target relative phases shown in Fig.~\ref{fig:cat}(b), demonstrating robust coherent control of both sideband populations and relative phases.

\subsection{Three-Sideband Quantum States}
\begin{figure*}
\includegraphics[width=1\linewidth]{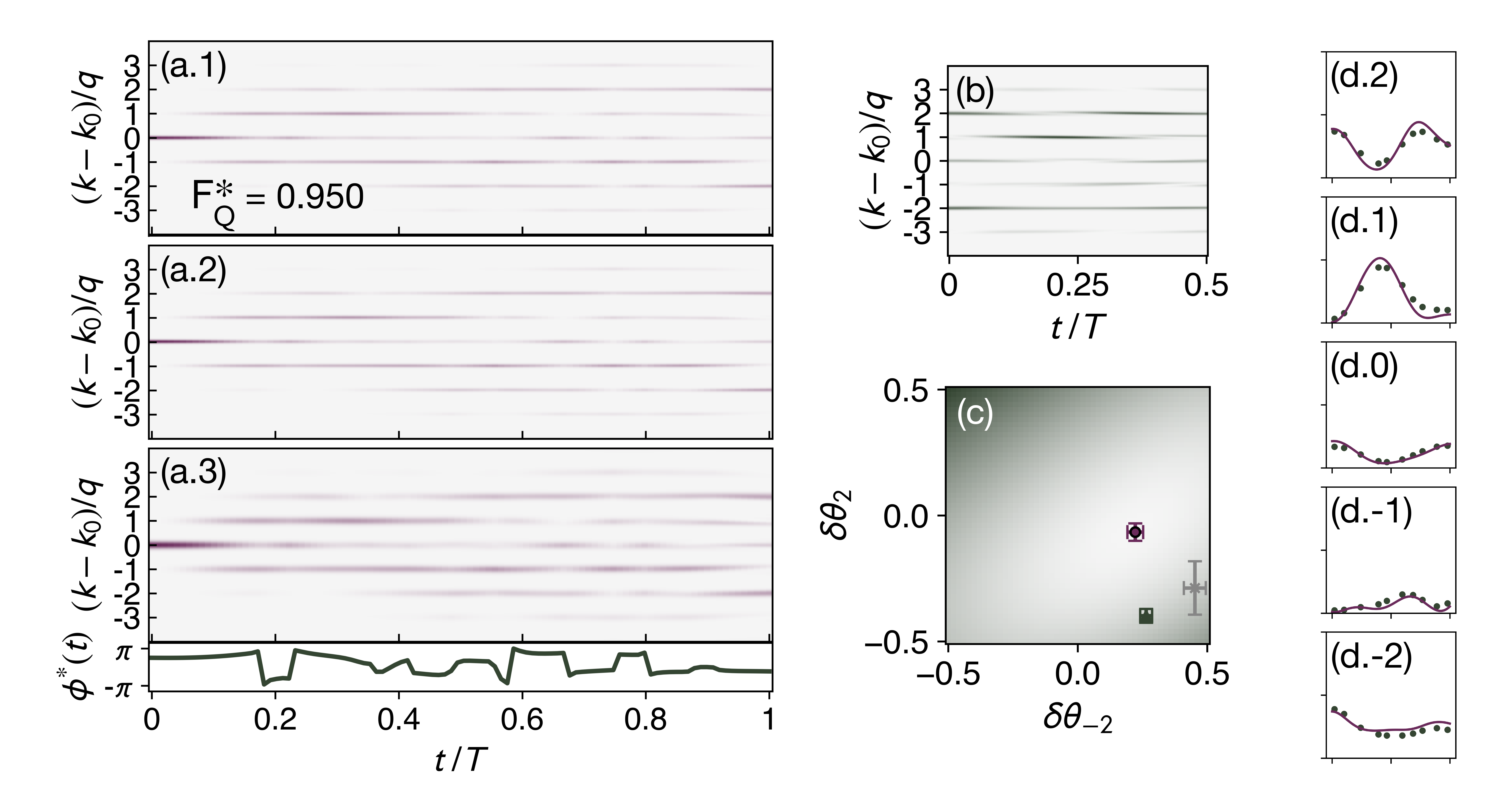}
\caption{\label{fig:three}Preparation of the three-sideband target state $|C_t\rangle = e^{-i\pi/2}|-2\rangle + |0\rangle + e^{i\pi/2}|2\rangle$ through single parameter control. (a.1-3) Optimized control field $\phi^*(t)$ and corresponding momentum space dynamics for (a.1) perfect phase matching, (a.2) detuned interaction $\delta k = 0.1q$, and (a.3) broader momentum space wavepacket with $\sigma_k=0.1q$. The ideal Floquet-Bloch optimization yields a quantum fidelity of $F_Q^*=0.950$. (b) Additional free evolution dynamics for the perfect phase matching case in (a.1) over a duration of $0.5T$ under a static control field $\phi(t)=0$. (c) Least-square-error landscape for reconstructing the relative phases between sidebands according to Eq.~\eqref{eq:interference} under perfect phase matching. The optimal fitting points corresponding to perfect phase matching (purple dot), detuned interaction (green triangle), and broader wavepacket dynamics (gray cross) are shown on the same landscape for illustrative purposes. (d.-2 to 2) Least square fits of the sideband populations used to reconstruct the relative phases for the perfect phase-matching case.}
\end{figure*}

We next consider a more complex target state consisting of a coherent superposition of three momentum sidebands,
\begin{equation}
|C_t\rangle = \frac{1}{\sqrt{3}}\left(e^{-i\pi/2}|-2\rangle + |0\rangle + e^{i\pi/2}|2\rangle\right).
\end{equation}
The ideal momentum space evolution and corresponding optimal control field $\phi^*(t)$ are shown in Fig.~\ref{fig:three}(a.1). We further investigate the robustness of the coherent control protocol under two experimentally relevant imperfections: detuned light-electron interaction and increased spatial localization of the electron wavepacket.

In the coupled-mode model~\eqref{eq:tbeq}, perfect phase matching is imposed through the condition $v_0q=\omega_L$ at the central momentum $p_0$. A finite detuning $\Delta = v_0q-\omega_L$ introduces an additional linear potential term in the coupled-mode dynamics,
\begin{equation}
\label{eq:detuned}
i\dot{c_n} = n\Delta c_n + n^2\epsilon c_n + \kappa c_{n+1} + \kappa^* c_{n-1}.
\end{equation}
As a result, the evolution deviates from the ideal resonant dynamics and leads to fidelity degradation when the detuning acts as an uncontrolled systematic error. To illustrate this effect, we shift the electron central momentum by $0.1q$ relative to $k_0$, corresponding to the detuned dynamics shown in Fig.~\ref{fig:three}(a.2).

We further consider a more spatially localized electron wavepacket by increasing the momentum space width to $\sigma_k=0.1q$. In this regime, the Floquet-Bloch approximation becomes less accurate, and the Gaussian sidebands exhibit noticeable distortion during the evolution [see Fig.~\ref{fig:three}(a.3)]. The underlying mechanism is analogous to the anomalous Bragg effect previously reported in reduced two-level descriptions of light-electron interactions.

To reconstruct the coherent superposition and evaluate the experimentally accessible quantum fidelity, we again evolve the electron under a static control field $\phi(t)=0$ following the optimized preparation stage. The ideal post-evolution dynamics are shown in Fig.~\ref{fig:three}(b). Similar to~\eqref{eq:gaussian_cat}, we reconstruct the TDSE output state $|\Psi_k(T)\rangle$ using the Gaussian ansatz for interference
\begin{eqnarray}
\label{eq:interference}
|\Psi\rangle &=& \mathcal{N}\left\{\sum_{n\neq\pm2} c_i \exp\left[-\frac{(k-k_0-nq)^2}{2\sigma_k^2}\right]\right.\nonumber\\
&+& c_{-2} e^{i\delta\theta_{-2}}\exp\left[-\frac{(k-k_0+2q)^2}{2\sigma_k^2}\right] \nonumber\\
&+&  \left.c_2 e^{i\delta\theta_{2}}\exp\left[-\frac{(k-k_0-2q)^2}{2\sigma_k^2}\right]\right\},
\end{eqnarray}
where $\delta\theta_{\pm2}$ denote independent phase perturbations on the two sidebands. The introduction of two fitting parameters enlarges the interferometric parameter space from one to two dimensions. Figure~\ref{fig:three}(c) shows the optimal estimates $\delta\theta_{\pm2}^*$ obtained through least square fitting of the measured population dynamics [see Fig.~\ref{fig:three}(d.-2 to 2)]. The reconstructed relative phases are then given by
\begin{equation}
\Delta\theta_{m,-2} = \text{arg}(c_{-2}/c_{0}) + \delta\theta_{-2}^*,~ \Delta\theta_{m,2} = \text{arg}(c_{2}/c_{0}) + \delta\theta_{2}^*.
\end{equation}
The reconstructed quantum fidelities for the three configurations are $F_Q = \{0.880, 0.813, 0.718\}$, demonstrating that the coherent control protocol remains robust against moderate detuning and wavepacket localization effects. These results further indicate that high fidelity quantum control requires precise phase matching and sufficiently narrow momentum space sidebands to maintain the validity of the Floquet-Bloch approximation. Notably, the detuning term can alternatively be interpreted as a synthetic linear potential in momentum space, analogous to a Wannier-Stark ladder in driven lattice systems. Under the gauge transformation $c_n(t)\rightarrow c_n(t)e^{-in\Delta t}$, the detuning can be absorbed into a time-dependent hopping phase, resulting in an effective phase shift $\phi(t)\rightarrow\phi(t)-\Delta t$. This observation suggests that part of the detuning-induced fidelity loss can be compensated through adaptive phase correction of the control field if the detuning error is known.

\section{Deterministic Sequential Control}
\label{sec:dsc}
From the single parameter control examples presented above, we conclude that ultrafast quantum control of free electron momentum spectra can be achieved through optimization of the optical phase $\phi(t)$. Nevertheless, several important limitations of the present framework should be emphasized.

First, the coupled-mode equation~\eqref{eq:tbeq} is derived under the approximation $k_n\approx k_0$, which leads to homogeneous hopping amplitudes between neighboring sidebands. This approximation gradually breaks down when targeting higher-order momentum states far from the phase matching point, where the momentum dependence of the light-electron coupling becomes non-negligible. As a result, the effective tight-binding description may no longer accurately reproduce the full dynamics of the minimal-coupling Hamiltonian.

Second, the introduction of a time-dependent phase $\phi(t)$ formally modifies the periodic structure underlying the Floquet-Bloch theory. In particular, the temporal variation of the phase contributes an additional instantaneous frequency shift through $\dot{\phi}(t)$, such that the effective driving frequency becomes $\omega_L+\dot{\phi}(t)$. The coupled-mode description therefore remains valid only when the phase modulation varies sufficiently slowly compared with the optical carrier frequency, i.e., $|\dot{\phi}(t)|\ll\omega_L$. Rapid phase modulation may induce corrections beyond the present Floquet approximation and modify the effective light-electron interaction. This could be problematic when we employ Pontryagin's maximum principle for optimization, resulting in bang-bang-like control field.

From the perspective of experimental implementation, preparation of distant momentum sidebands generally requires either larger hopping amplitudes $\kappa$ or longer interaction times. For low-energy electrons, achieving sufficiently large $\kappa$ may demand extremely strong optical fields, which can be challenging experimentally. Moreover, increasing $\kappa$ drives the system toward the Raman-Nath regime, where diffraction spreads across many momentum sidebands simultaneously. In this regime, the target population becomes distributed over a broad momentum range, resulting in a flatter optimization landscape and significantly reduced gradient signals for specific target states. Consequently, numerical optimization becomes increasingly difficult for high order sideband preparation.

From the optimization perspective, the present gradient-based approach also depends sensitively on the initial ansatz and learning rate, and may become trapped in local optima. In practice, obtaining satisfactory control fields often requires substantial tuning of hyperparameters or the incorporation of additional heuristic optimization strategies on a case-by-case basis. These considerations naturally motivate the search for deterministic control protocols that avoid large-scale numerical optimization altogether.

\subsection{Dynamical Phase Matching}

As discussed in the case of the three-sideband state, violation of the phase-matching condition $v_0 q = \omega_L$ introduces an additional linear detuning term $\Delta n$ in the coupled-mode equation~\eqref{eq:detuned}, where $\Delta = v_0q-\omega_L$. Combined with the quadratic on-site dispersion $n^2\epsilon$, the effective resonance condition is shifted to $n_c = -\frac{\Delta}{2\epsilon}$, corresponding to a displaced parabolic potential in synthetic momentum space.

On the other hand, the coupled-mode equation~\eqref{eq:tbeq} shows that higher-order sidebands can be strongly suppressed in the Bragg regime $Q\gg1$, allowing the dynamics to be reduced to an effective two-level system. Choosing the resonance center at $n=\pm1/2$ yields
\begin{eqnarray}
i\dot{c}_{\frac{1}{2}} = \frac{\epsilon}{4} c_{\frac{1}{2}} + \kappa^* c_{-\frac{1}{2}},\nonumber\\
i\dot{c}_{-\frac{1}{2}} = \frac{\epsilon}{4} c_{-\frac{1}{2}} + \kappa c_{\frac{1}{2}},
\end{eqnarray}
where the resonant coupling remains centered around $n=0$. More generally, neighboring sidebands can be selectively brought into resonance by tuning the laser frequency according to
\begin{equation}
\delta\omega_j = 2\epsilon\left(j+\frac{1}{2}\right),
\end{equation}
thereby realizing an effective two-level dynamics between the states $|j\rangle$ and $|j+1\rangle$. This observation motivates a deterministic control protocol for coherently preparing arbitrary states within a finite momentum space manifold through sequential resonant population transfer.

The protocol is implemented by passing the electron through a sequence of $N-1$ independent interaction zones, each driven by a tailored optical field. Throughout the protocol, the system operates in the deep Bragg regime, where the light-electron coupling strength remains small compared with the quadratic on-site energy scale. Under the condition $Q\gg1$, the interaction becomes highly frequency selective. By tuning the laser frequency in the $j$-th interaction zone as $\omega_L\rightarrow\omega_L+\delta\omega_j$, the resonance condition is shifted such that the momentum states $|j\rangle$ and $|j+1\rangle$ become resonantly coupled. At the same time, transitions to neighboring off-resonant states such as $|j-1\rangle$ and $|j+2\rangle$ remain strongly suppressed. The dynamics within each interaction zone can therefore be approximated as an isolated two-level system embedded within the larger synthetic momentum lattice.

The ideal evolution between the amplitudes $c_j$ and $c_{j+1}$ is described by the $\mathrm{SU}(2)$ rotation
\begin{equation}
\label{eq:tls}
U_j(\Theta_j, \phi_j) = \begin{pmatrix} \cos(\Theta_j/2) & -i e^{-i\phi_j} \sin(\Theta_j/2) \\ -i e^{i\phi_j} \sin(\Theta_j/2) & \cos(\Theta_j/2) \end{pmatrix},
\end{equation}
where $\Theta_j$ denotes the pulse area controlling the population transfer and $\phi_j$ is the optical phase encoded in the complex hopping amplitude $\kappa$.

If the dynamics consisted solely of sequential $\mathrm{SU}(2)$ rotations, the state-preparation protocol would be straightforward. In practice, however, all populated sidebands continuously accumulate dynamical phases due to the quadratic on-site dispersion during the finite interaction time $t_j$. While the resonant pulse selectively drives the states $|j\rangle$ and $|j+1\rangle$, previously prepared sidebands continue to acquire additional phases determined by their detuning from the instantaneous resonance condition according to Eq.~\eqref{eq:detuned}. Precise quantum state preparation therefore requires the optical phases $\phi_j$ to pre-compensate these deterministic dynamical phase accumulations throughout the sequential control protocol.

\subsection{Deterministic Inverse Engineering}

As in the single parameter control protocol, our objective is to prepare an arbitrary target electron state consisting of a coherent superposition of $N$ momentum sidebands,
\begin{equation}
|C_t\rangle = \sum_{j=0}^{N-1}c_j|j\rangle.
\end{equation}
The protocol is implemented by sequentially passing the electron through $N-1$ independent light-matter interaction zones. To determine the required pulse areas $\Theta_j$ and optical phases $\phi_j$ for each zone, we employ a deterministic backward-propagation algorithm. Starting from the target state $|C_t\rangle$, the procedure iterates backward from the highest addressed sideband transition to the lowest one.

In the $j$-th interaction zone, the resonant pulse is designed to coherently transfer all remaining population from the sideband $|j+1\rangle$ into $|j\rangle$. Under the ideal two-level approximation, the required rotation is determined from the complex amplitude ratio between the two adjacent sidebands,
\begin{equation}
r = i \frac{c_{j+1}}{c_j}.
\end{equation}
The corresponding pulse area and optical phase are then given by
\begin{equation}
\Theta_j = 2 \arctan(|r|),~\phi_j = \text{arg}(r).
\end{equation}
With this choice, the $\mathrm{SU}(2)$ rotation destructively eliminates the population in $|j+1\rangle$ during the backward-propagation step.

However, the evolution cannot be described solely by applying the inverse two-level rotation $U_j^\dag$~\eqref{eq:tls} to the selected sidebands. During the finite interaction time, all populated sidebands continue to accumulate dynamical phases according to the detuned coupled-mode equation~\eqref{eq:detuned}. The backward-propagation step must therefore be implemented through full time-reversed evolution under the detuned dynamics rather than an isolated two-level transformation. The pulse duration in each interaction zone is determined by
\begin{equation}
t_{p,j} = \frac{\Theta_j}{2|\kappa|}=\frac{\Theta_j Q}{\epsilon}.
\end{equation}

Once the iterative procedure reaches $j=0$, the backward evolution maps the target state entirely onto the initial state $|0\rangle$. The complete set of pulse areas $\Theta_j$ and optical phases $\phi_j$ thereby provides an exact deterministic protocol for preparing the desired target state through sequential resonant control under different Klein-Cook parameters $Q$.

\begin{figure}
\includegraphics[width=1\linewidth]{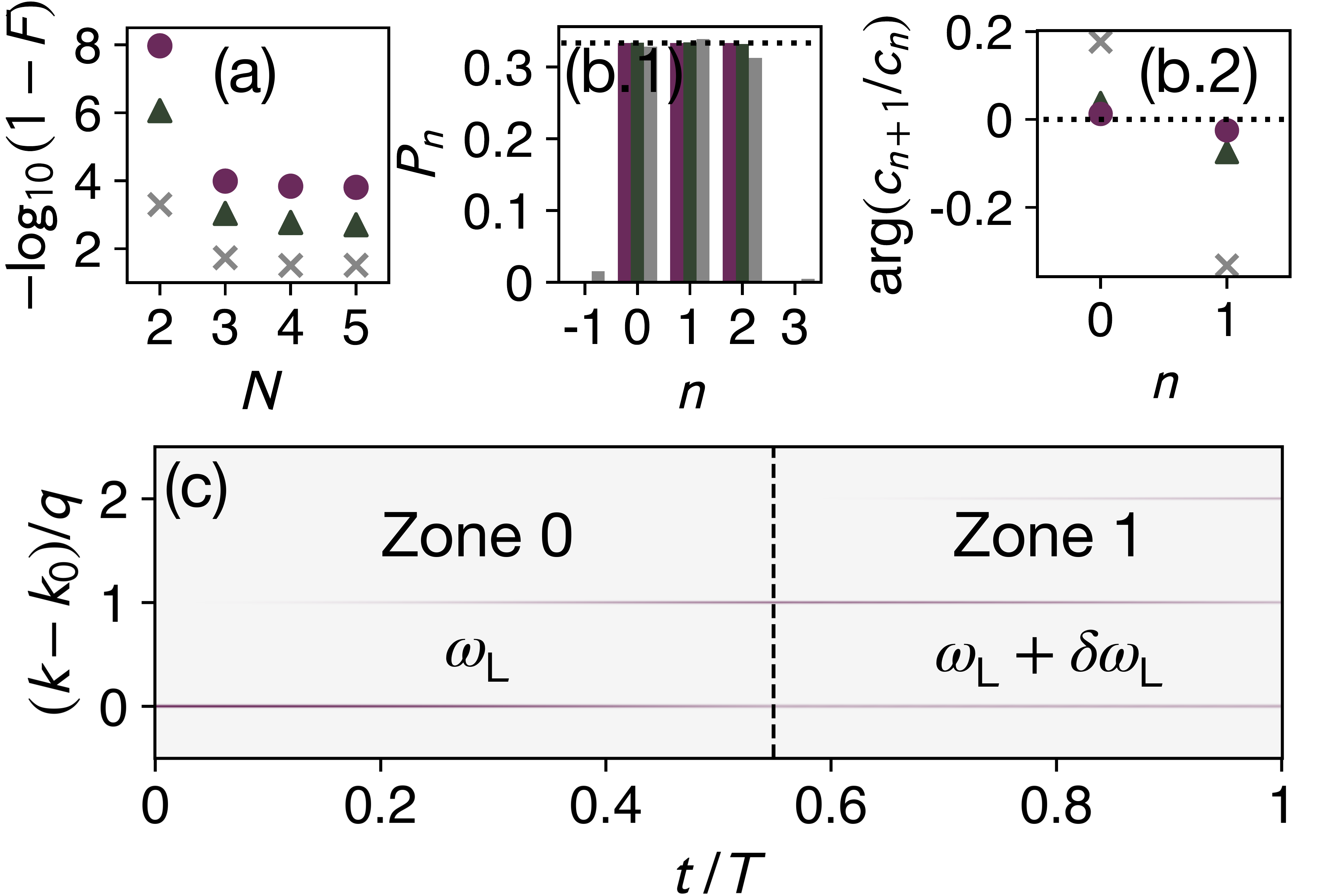}
\caption{\label{fig:deterministic} Deterministic preparation of equal-weight momentum space superposition states with zero relative phases. (a) Quantum fidelities for preparing superposition states spanning $N$ sidebands through sequential resonant control. Purple dots, green triangles, and gray crosses correspond to Klein-Cook parameters $Q=30$, $Q=10$, and $Q=2$, respectively. (b.1-2) Final sideband populations and relative phases between neighboring sidebands for the corresponding values of $Q$. (c) momentum space dynamics for the preparation protocol with $\{Q=10,~N=3\}$.}
\end{figure}

\subsection{Sequential Control Performance}

We now evaluate the performance of the deterministic control protocol by preparing equal-weight superposition states spanning $N$ momentum sidebands with relative phases set to zero. Figure~\ref{fig:deterministic}(a) compares the achieved quantum fidelities for different Klein-Cook parameters ranging from the deep Bragg regime $(Q=30)$, where the two-level approximation is well satisfied, to regimes with stronger multilevel coupling $(Q=2)$ that exhibit noticeable leakage into off-resonant sidebands. Remarkably, the protocol maintains high fidelities even beyond the idealized two-level limit, demonstrating the robustness of the sequential resonant-control strategy.

For the specific case of $N=3$, Fig.~\ref{fig:deterministic}(b.1) shows the final sideband populations, while Fig.~\ref{fig:deterministic}(b.2) presents the reconstructed relative phases. The results confirm that the deterministic protocol enables high-precision preparation of coherent momentum space superposition states.

Figure~\ref{fig:deterministic}(c) illustrates the full momentum space dynamics as the free electron sequentially traverses two interaction zones with independently tuned laser frequencies for dynamical phase matching, corresponding to the case $\{Q=10, N=3\}$ in Fig.~\ref{fig:deterministic}(a). Compared with the single parameter optimal control protocol, operation in the Bragg regime requires substantially longer interaction times in order to maintain frequency-selective coupling between neighboring sidebands. Consequently, a narrower initial momentum distribution is required to suppress wavepacket distortion induced by free electron dispersion during the evolution.

In the numerical simulations, the Klein-Cook parameter is fixed to $Q=10$ throughout the full control sequence. Due to the dynamical phase matching and gradual acceleration of the electron, the relativistic factor $\gamma$ changes slightly during the evolution, leading to weak variations in both the on-site energy $\epsilon=2.275\times10^{-2}~\mathrm{1/fs}$ and hopping amplitude $\kappa=1.138\times10^{-3}~\mathrm{1/fs}$. These corrections remain negligible compared with other approximations in the model and experimentally unavoidable systematic errors. The pulse durations in the two interaction zones are $t_{p,0}=839.48~\mathrm{fs}$ and $t_{p,1}=690.40~\mathrm{fs}$, respectively. To suppress dispersion-induced dephasing throughout the long interaction sequence, we choose a narrow momentum space width $\sigma_k=0.01q$.

To experimentally extract the relative phases between sidebands, one may employ a procedure analogous to the interferometric reconstruction method discussed previously by tuning the phase-matching condition midway between neighboring sidebands and performing deterministic Ramsey-type interference measurements. Simulating the full minimal-coupling Hamiltonian through the time-dependent Schr\"odinger equation yields a final quantum fidelity of $F_Q=0.998$, in excellent agreement with the theoretical prediction.

As a reference, a more localized initial wavepacket with $\sigma_k=0.03q$ still achieves a high fidelity of $F_Q=0.991$. However, the fidelity rapidly decreases to $F_Q=0.085$ for $\sigma_k=0.05q$, primarily due to wavepacket distortion and dephasing induced by free electron dispersion. We further observe that the relative phases between lower sidebands are highly sensitive to the optical phases in the early stages of the control sequence. For general target states, even small perturbations of the Klein-Cook parameter $Q$ can substantially modify the required laser phases for subsequent interaction zones. Accurate coherent state preparation therefore demands extremely precise phase calibration throughout the sequential protocol. In contrast, the final sideband populations remain comparatively robust against such phase errors provided that the electron wavepacket remains sufficiently narrow in momentum space. This robustness makes population-only state engineering considerably more experimentally accessible, as discussed in Section~\ref{sec:pop}.

The two control protocols developed in this work also reveal a fundamental tradeoff between speed, controllability, and dynamical selectivity in synthetic momentum space engineering. The optimal control approach operates naturally in the intermediate or Raman-Nath regimes, where multilevel diffraction and coherent interference enable ultrafast state preparation through globally optimized dynamics. In contrast, the deterministic sequential protocol relies on the frequency selectivity of the Bragg regime, reducing the dynamics to controllable resonant two-level transitions at the cost of substantially longer interaction times and increased sensitivity to wavepacket dispersion. These complementary regimes therefore provide distinct routes toward free electron quantum control: fast interference-based manipulation through global optimization, and analytically structured state synthesis through sequential resonant engineering.

\section{Conclusion and outlook}

In this work, we developed coherent control protocols for free electron quantum states in synthetic momentum space generated through light-electron interactions at optical gratings. Using a Floquet-Bloch tight-binding description derived from the minimal-coupling Hamiltonian, we demonstrated both programmable optimal control based on Pontryagin's maximum principle and deterministic sequential control based on dynamical phase matching in the Bragg regime. These approaches enable high-fidelity preparation of momentum-sideband populations and coherent superposition states with programmable relative phases, in good agreement with numerical simulations of the time-dependent Schr\"odinger equation. Our results establish synthetic momentum lattices as a versatile framework for coherent free electron quantum state engineering beyond conventional mechanisms.

The present work opens several promising directions for future exploration. Extending the light-electron interaction toward multimode optical driving could enable programmable synthetic momentum lattices with long-range hopping and higher-dimensional synthetic structures~\cite{yuan2018synthetic}. In such systems, independently controlled hopping amplitudes and phases may provide access to Floquet gauge fields, engineered band structures, and topological momentum space dynamics~\cite{ozawa2019topological} for free electrons. Beyond the minimal coupling light-electron interaction considered here, incorporating nonlinear optical processes may further induce exotic electron dynamics~\cite{konecna2020nanoscale} with multi-photon transitions, parametrically driven hopping processes, and nonperturbative effects on synthetic lattices. These extensions could considerably enrich the accessible nonequilibrium quantum dynamics of free electron in synthetic dimensions.

The ability to coherently engineer amplitudes and relative phases between momentum sidebands also naturally suggests applications in momentum space interferometry and ultrafast quantum metrology. Since the relative phases between sidebands directly determine the spatiotemporal structure of the electron wavefunction, programmable momentum space synthesis enables controllable shaping of structured electron pulses through coherent Fourier engineering. In particular, tailored superpositions of momentum sidebands may enable highly localized or superoscillatory free electron wavepackets~\cite{berry2019roadmap,remez2017superoscillating} with nontrivial temporal and spatial interference structures. More broadly, coherent control of free electron synthetic momentum states may provide a versatile platform for programmable quantum transport, synthetic quantum dynamics, and analog quantum simulation with ultrafast electron beams.

\section*{Code availability}
All codes for optimization and simulation are available in this repository~\cite{github}.

\section*{Acknowledgement}
A. A., M. L., and M. S. thank support from the Basque Government BasQ initiative under the Q-STREAM project. They also acknowledge support from OpenSuperQ+100 (Grant No. 101113946) of the EU Flagship on Quantum Technologies, from Project Grant No. PID2024-156808NB-I00 and Spanish Ram´on y Cajal Grant No. RYC-2020-030503-I funded by MICIU/AEI/10.13039/501100011033 and by "ERDF A way of making Europe" and "ERDF Invest in your Future", and from the Spanish Ministry for Digital Transformation and of Civil Service of the Spanish Government through the QUANTUM ENIA project call Quantum Spain, and by the EU through the Recovery, Transformation and Resilience Plan-Next Generation EU within the framework of the Digital Spain 2026 Agenda.Y.P. was supported by the NSFC (No. 2023X0201-417-03) and the start-up funding from ShanghaiTech University. Y. D. thanks the European Commission for a Marie Curie PF grant (No. 101204580 FELQO).

\end{document}